\documentclass[preprint, flushrt]{aastex}

\newcommand{\etal}{\textrm{et al. }}

\newcommand{\eg}{\textrm{e.g. }}

\newcommand{\Angstrom}  {\,{\rm \AA}}
\newcommand{\um}        {\,\mu{\rm m}}
\newcommand{\kpc}        {\,{\rm kpc}}
\newcommand{\Ha}        {\,{\rm {H\alpha}}}
\newcommand{\Hb}        {\,{\rm {H\beta}}}

\catcode`\@=11 
\def\lsim{\mathrel{\mathpalette\@versim<}}
\def\gsim{\mathrel{\mathpalette\@versim>}}
\def\@versim#1#2{\vcenter{\offinterlineskip
        \ialign{$\m@th#1\hfil##\hfil$\crcr#2\crcr\sim\crcr } }}

\begin{document}

\title{Active Galactic Nuclei in the Sloan Digital Sky Survey: I. Sample Selection}
\author{
Lei~Hao\altaffilmark{1,2}, 
Michael~A.~Strauss\altaffilmark{1}, 
Christy~A.~Tremonti\altaffilmark{3}, 
David~J.~Schlegel\altaffilmark{1},
Timothy~M.~Heckman\altaffilmark{4}, 
Guinevere~Kauffmann\altaffilmark{5}, 
Michael~R.~Blanton\altaffilmark{6}, 
Xiaohui~Fan\altaffilmark{3},
James~E.~Gunn\altaffilmark{1},
Patrick~B.~Hall\altaffilmark{1}, 
\v{Z}eljko~Ivezi\'{c}\altaffilmark{1}, 
Gillian~R.~Knapp\altaffilmark{1}, 
Julian~H.~Krolik\altaffilmark{4},
Robert~H.~Lupton\altaffilmark{1},
Gordon~T.~Richards\altaffilmark{1},
Donald~P.~Schneider\altaffilmark{7}, 
Iskra~V.~Strateva\altaffilmark{1}, 
Nadia~L.~Zakamska\altaffilmark{1},
J.~Brinkmann\altaffilmark{8},
Robert~J.~Brunner\altaffilmark{9},
Gyula~P.~Szokoly\altaffilmark{5}}
\altaffiltext{1}{Princeton University Observatory, Princeton, NJ 08544}
\altaffiltext{2}{Current address: Astronomy Department, Cornell University, Ithaca, NY 14853; haol@isc.astro.cornell.edu}
\altaffiltext{3}{Steward Observatory, University of Arizona, 933 North Cherry Avenue, Tucson, AZ 85721}
\altaffiltext{4}{Department of Physics and Astronomy, Johns Hopkins University, 3400 North Charles Street, Baltimore, MD 21218} 
\altaffiltext{5}{Max-Planck Institut f\"{u}r Astrophysik, D-85748 Garching, Germany}
\altaffiltext{6}{Center for Cosmology and Particle Physics, Department of Physics, New York University, 4 Washington Place, New York, NY 10003}
\altaffiltext{7}{Department of Astronomy and Astrophysics, Pennsylvania State University, University Park, PA 16802}
\altaffiltext{8}{Apache Point Observatory, P.O. Box 59, Sunspot, NM 88349-0059.}
\altaffiltext{9}{Department of Astronomy and National Center for Supercomputer Applications, University of Illinois, 1002 West Green Street, Urbana, IL 61801.}
\slugcomment{\textit{\today}}

\begin{abstract}
We have compiled a large sample of low-redshift active galactic nuclei
(AGN) identified via their emission line characteristics from the
spectroscopic data of the Sloan Digital Sky Survey. Since emission
lines are often contaminated by stellar absorption lines, we developed
an objective and efficient method of subtracting the stellar continuum
from every galaxy spectrum before making emission line
measurements. The distribution of the measured H$\alpha$ Full Width at
Half Maxima values of emission line galaxies is strongly bimodal, with
two populations separated at about 1,200km s$^{-1}$. This feature
provides a natural separation between narrow-line and broad-line
AGN. The narrow-line AGN are identified using standard emission line
ratio diagnostic diagrams. 1,317 broad-line and 3,074 narrow-line AGN
are identified from about 100,000 galaxy spectra selected over 1151
square degrees. This sample is used in a companion paper to determine
the emission-line luminosity function of AGN.
\end{abstract}


\section{Introduction}

Ever since the definition of Seyfert galaxies (Seyfert 1943) and the
first recognition of quasars (Schmidt 1963), astronomers have put
enormous effort into compiling large samples of active galactic nuclei
(AGN, in this paper, ``AGN'' refers to active galactic nuclei at all
luminosities, including quasars) and trying to understand the physics
that powers them. This is not easy, since luminous AGN comprise only a
few percent of normal galaxies. Based on the distinctive
characteristics of AGN, different methods have been developed to
search for AGN in various wavebands.

In the optical, AGN show different colors from stars and normal
galaxies, especially at high luminosity. Many surveys have used color
selection (\eg Schmidt \& Green 1983; Boyle \etal 1990). In
particular, the Sloan Digital Sky Survey (SDSS) (York \etal 2000) uses
optical colors to identify quasar candidates, which are then observed
spectroscopically (Richards \etal 2002). The color selection is very
efficient but it requires that the optical luminosity of an AGN be at
least comparable to the luminosity of its host galaxy for the color to
be distinctive, and thus the color selection systematically misses
less luminous AGN at low redshift.

AGN also show strong optical and ultraviolet emission lines. Broadly
speaking, AGN can be classified into two types: broad-line and
narrow-line AGN. The former show broad permitted emission lines, with
Full Width at Half Maxima (FWHM) of several thousand km s$^{-1}$,
while in narrow-line AGN, both the permitted and forbidden emission
lines are narrow, with FWHMs $\sim$ 500 km s$^{-1}$. This is
comparable with emission lines in normal star-forming galaxies, but
emission lines in narrow-line AGN have considerably greater ionization
range. In particular, both high-ionization lines such as
[NeIII]$\lambda 3869$, [Ne V]$\lambda 3426$ and [OIII]$\lambda 5007$,
and low-ionization lines such as [OI]$\lambda 6300$ and
[NI]$\lambda5200$, are stronger in narrow-line AGN than in normal
starforming galaxies. Based on this feature, narrow-line AGN can be
identified by their distinctive emission line ratios.

The first line ratio diagram was introduced by Baldwin, Phillips \&
Terlevich (1982), who suggested that AGN generically have greater
[OIII]$\lambda 5007$/H$\beta$ (the flux ratio of [OIII]$\lambda$5007
to H$\beta$) than do galaxies whose emission lines are due to stellar
processes. Veilleux \& Osterbrock (1987) developed their idea and used
diagnostic diagrams that consist of combinations of four line ratios:
[OIII]$\lambda$5007/H$\beta$, [NII]$\lambda 6584$/H$\alpha$,
[OI]$\lambda6300$/H$\alpha$ and [SII]$\lambda\lambda
6716,31$/H$\alpha$, and further developed a semi-empirical line on the
diagrams to separate AGN and starburst galaxies. The emission-line
pairs are chosen specifically so that the two lines in a ratio are at
nearly identical wavelengths, therefore reddening and
spectrophotometric uncertainties are not big effects. These diagnostic
diagrams have been used ever since as a standard to identify
narrow-line AGN.

Kewley \etal (2001) developed a set of theoretical separation lines
for AGN and starforming galaxies on the diagnostic diagrams. By
constructing a detailed continuous starburst model with large
realistic metallicity and ionization parameter ranges, they found that
the model folds on the diagnostic diagrams and there exist upper
limits for starforming galaxies. These upper limits can be used to
separate AGN and starforming galaxies and they can be fitted to a
simple rectangular hyperbolic shape:
\begin{eqnarray}
 \log\left(\frac{\mathrm{[OIII]}\lambda 5007}{\mathrm{H\beta}}\right) = \frac{0.61}{\log(\mathrm{[NII]/H\alpha})-0.47}+1.19 \nonumber \\
 \log\left(\frac{\mathrm{[OIII]}\lambda 5007}{\mathrm{H\beta}}\right) = \frac{0.72}{\log(\mathrm{[SII]/H\alpha})-0.32}+1.30 \\
 \log\left(\frac{\mathrm{[OIII]}\lambda 5007}{\mathrm{H\beta}}\right) = \frac{0.73}{\log(\mathrm{[OI]/H\alpha})+0.59}+1.33 \nonumber
\label{eq:kewleyline}
\end{eqnarray}

These theoretical separation lines between AGN and starforming
galaxies are widely accepted to use to identify narrow-line AGN in the
diagnostic diagrams.

Kauffmann \etal (2003a), on the other hand, when studying
host galaxy properties of narrow-line AGN in the SDSS, proposed an empirical and more lenient cut to identify AGN:
\begin{equation}
 \log\left(\frac{\mathrm{[OIII]}\lambda 5007}{\mathrm{H\beta}}\right)
 \ge \frac{0.61}{\log(\mathrm{[NII]/H\alpha})-0.05}+1.3
\label{eq:kauff}
\end{equation}
This criterion will select many more galaxies as AGN than Kewley's
criteria. In this paper, we apply both criteria and discuss their
differences in detail.

There are several spectroscopic galaxy surveys from which people have
tried to select AGN based on their emission lines. One is the CfA
redshift survey (Davis, Huchra \& Latham 1983; Huchra \etal
1983). Spectra of about 2,400 galaxies were taken to study their large
scale distribution. Huchra, Wyatt \& Davis (1982) used these spectra
to select AGN by identifying emission lines indicative of nonstellar
activity. As a result they found roughly 50 Seyfert galaxies, divided
approximately equally between Seyfert 1 and Seyfert 2 galaxies, making
the local Seyfert fraction $\sim 2\%$ (Huchra \& Burg 1992).

Another example is the Revised Shapley-Ames catalog of bright galaxies
(Sandage \& Tammann 1981). Among its $\sim$1,300 galaxies, about 50
are Seyfert galaxies of luminosity comparable to those found in the
CfA Survey. Ho \etal (1997a) took uniform high signal to noise ratio
(S/N) spectra of 486 galaxies using very small apertures centered on
the nuclei. 418 galaxies were found to contain emission-line nuclei,
of which 206 are star-forming, and 211 show AGN activity. This
demonstrates that low-luminosity AGN activity in the local universe is
extremely common, as first discussed by Phillips, Charles, \& Baldwin
(1983).

Hall \etal (2000) systematically selected high redshift AGN based
either on their broad emission lines or narrow [NeV] emission lines
from the Canadian Network for Observational Cosmology field galaxy
redshift survey (CNOC2; Yee \etal 2000). They found 47 confirmed and
14 candidate AGN in the redshift range $0.27 \le z \le 4.67$.

The SDSS has opened a new window to the study of AGN by providing a
huge number of high quality spectra of galaxies. There have been
several studies identifying AGN spectroscopically from different
subsamples of SDSS galaxies (Kauffmann \etal 2003a; Miller \etal
2003). Both studies have focused on narrow-line AGN identified from
the diagnostic diagrams. Miller \etal (2003) found AGN signatures in
at least 20\% of 4,921 galaxies in the redshift range $0.05<z<0.095$,
and studied the environment of these AGN. Kauffmann \etal (2003a)
selected 22,623 narrow-line AGN from a parent sample of 122,808
galaxies and studied their host galaxy properties. The AGN detection
rate, however, depends very much on the AGN selection criteria used
and many other details in defining the sample. In this paper, we apply
a systematic search for both broad and narrow-line AGN in a
well-defined sky area in the redshift range $0<z<0.33$. The framework
of our narrow-line AGN selection is similar to Kauffmann \etal
(2003a), but we differ in many details.
 
In section 2, we give a brief overview of the SDSS, focusing on those
aspects most relevant to our study. $\S3$ introduces the parent sample
from which we will select our AGN. In $\S4$, we will discuss the
subtraction of stellar absorption lines from the spectra. The emission
line measurements are discussed in $\S5$. In $\S6$, the selected AGN
are presented and discussed. $\S7$ gives a clean AGN sample and we
summarize in $\S8$.

\section{The Sloan Digital Sky Survey}
 The Sloan Digital Sky Survey (York \etal 2000) is an imaging and
spectroscopic survey that will eventually cover approximately
one-quarter of the Celestial Sphere and collect spectra of $\sim 10^6$
galaxies and $10^5$ quasars.  It uses a dedicated 2.5m telescope at
Apache Point, New Mexico, with a $3$ degree field, and a mosaic CCD
camera and two fiber-fed double spectrographs to carry out the imaging
and spectroscopic surveys respectively. A separate 20$\arcsec$
photometric telescope is used for photometric calibration (Smith \etal
2002; Hogg \etal 2001). The imaging camera (Gunn \etal 1998) consists
of a mosaic of 30 imaging CCDs with $24\um$ pixels subtending
0.396$\arcsec$ on the sky. The sky is observed through five broad-band
filters $(u, g, r, i, z)$ (Fukugita \etal 1996; Stoughton \etal 2002)
covering the entire optical band from the atmospheric cutoff in the
blue to the sensitivity limit of silicon CCDs in the red. The imaging
is done in drift-scan mode and the total integration time per filter
is 54.1s. The 50\% completeness limits for point sources are 22.5,
23.2, 22.6, 21.9 and 20.8 magnitudes respectively and the photometric
calibration is reproducible to 3\%, 2\%, 2\%, 2\% and 3\% for the five
bandpasses, respectively.  The image data are processed by a series of
automated pipelines (Lupton \etal 2001; Stoughton \etal 2002; Pier
\etal 2003), which make various measurements of the flux of each
detected object.

The SDSS is obtaining spectra of complete samples of three categories
of objects: Galaxies, Luminous Red Galaxies and Quasars. These
spectroscopic targets are selected from the imaging data via various
target selection criteria.  Galaxy target selection is discussed in
Strauss \etal (2002). Briefly speaking, galaxies are separated from
stars by morphology. The magnitude limit cut for the galaxy sample was
changed several times during commissioning, but is currently
$r=17.77$, where $r$ represents the $r$ band Petrosian magnitude.  All
magnitudes are corrected for extinction following Schlegel \etal
(1998). These objects are the sample from which we will select AGN.

Quasar target selection (Richards \etal 2002) is based on the
nonstellar colors of quasars and matching unresolved sources to the
FIRST radio catalog.  Luminous Red Galaxies are selected (Eisenstein
\etal 2001) by a variant of the photometric redshift method, aiming to
have a uniform, approximately volume-limited sample of highly luminous
objects with the reddest colors in the rest frame to $z=0.5$. In this
paper, we will not discuss AGN selected from these two samples, unless
they also satisfy the galaxy sample magnitude cut $r=17.77$.

The SDSS spectra are taken with two fiber-fed spectrographs, covering
the wavelength range 3800-9200 $\Angstrom$ over 4098 pixels. Each
plate can hold 640 fibers, with a fixed aperture of 3$\arcsec$. The
plates are positioned by a tiling algorithm (Blanton \etal 2003) and
fibers are assigned to targets. Galaxies are among the tiled targets
that have the highest priority of having their spectra taken. The
finite diameter of the fiber cladding prevents fibers on any given
plate from being placed closer than 55$\arcsec$ apart. The resolution
$\lambda/\Delta \lambda$ varies between 1850 and 2200. The relative
spectrophotometry is accurate to about 20\%. Each spectrum is
accompanied by an estimated error per pixel, based on photon
statistics and the amplitude of sky residuals. The typical S/N for
galaxy spectra at the sample limit is 16/pixel.

The spectroscopic data are reduced through the spectroscopic
pipelines, spectro2d and specBS. Spectro2d reduces the 2-dimensional
spectrograms produced by the spectrographs to flux- and
wavelength-calibrated spectra. SpecBS is different from the SDSS
official pipeline, it determines classifications and redshifts via a
$\chi^2$ fit to the spectrum in question with a series of rest-frame
star, galaxy and quasar templates. The basic technique is described by
Glazebrook \etal (1998) and Bromley \etal (1998).

Even though specBS has made measurements on emission lines by fitting
a single Gaussian at positions of each expected emission line, we
carry out our own fits in our study. The main reason is that a single
Gaussian fit is not adequate to model some AGN having both broad and
narrow emission lines. The emission lines are measured directly from
calibrated spectra, as we describe in detail in $\S 5$.

\section{Parent Sample}
In order to do statistical studies of AGN, it makes sense to start
with a complete sample of objects within a certain well defined area
from the SDSS. We start from 129,625 target objects complete in 1151
square degrees. This is about 1/2 of the spectra available in the SDSS
Second Data Release (Abazajian \etal 2004). Among these objects, there
are 98,684 galaxies targeted by the main galaxy target algorithm
(Strauss \etal 2002; $\S 2$) and 17,972 quasar target objects
(Richards \etal 2002; $\S 2$). In this paper, we will focus on the
galaxy target objects. However, there are 2057 extragalactic objects
selected as quasar targets with $r(\mathrm{Petrosian})\le 17.77$,
which were not targeted as galaxies because they are unresolved, and
we include them in our galaxy sample. Among these 100,741 galaxies,
most of them (99,990) have redshift $z<0.33$, guaranteeing that the
$\Ha$ emission line lies in our spectral coverage.  $\Ha$ is a very
useful emission line in identifying AGN
(Equation~\ref{eq:kewleyline}), and we limit our galaxy sample to
these 99,990 objects.

\section{Stellar Subtraction}

AGN are identified by their emission line characteristics. As
described in $\S 2$, the SDSS galaxy spectra are taken through a fixed
$3\arcsec$ aperture, which is large enough to let through not only the
light from the nucleus but also substantial amounts of stellar light
from the host galaxy. For example, at the median redshift of the
sample ($z=0.1$), a $3\arcsec$ aperture subtends about
$4h^{-1}\kpc$. Moreover, galaxies with higher redshift will have a
larger host galaxy component in the observed spectra. Thus the nuclear
emission lines are often contaminated by the stellar absorption lines
of the host galaxy. For weak AGN, this contamination can be so severe
that the interesting emission lines are completely submerged in the
absorption lines. Thus before considering AGN selection, we have to
develop a technique to properly remove the stellar absorption lines.

The basic idea of stellar subtraction is to build a library of stellar
absorption line spectra templates, and use them as building blocks to
simulate the stellar spectrum of the object in question. The library
needs to be complete in the sense that it contains enough information
on various absorption features to be able to simulate the stellar
components of various galaxies with widely spread metallicities, ages
and velocity dispersions. The library is typically composed of star
spectra generated from a population synthesis code (Bica 1988; Saraiva
\etal 2001, Kauffmann \etal 2003b) or direct observational spectra of
absorption line galaxies (Ho \etal 1997a) or stars (Engelbracht \etal
1998). In this paper the library is constructed by applying the
Principal Component Analysis (PCA) technique (e.g Connolly \etal 1995;
Lahav \etal 1996; Bromley \etal 1998; Eisenstein \etal 2003; Yip \etal
2004) to a sample of pure absorption-line galaxies.

The advantages of building stellar absorption templates via this
method are two-fold: only the first few eigenspectra are significant,
so we can limit the size of the library without losing much useful
information. Moreover, the eigenspectra are orthogonal to each other,
resulting in a unique solution to the stellar subtraction fit using
these templates.

\subsection{Preparing the Absorption Line Galaxy Sample}
We wish to identify a sample of pure absorption-line galaxies. In
practice, we require that the $\Ha$ equivalent width EW($\Ha$)$<$0
(positive EW corresponds to emission lines), and that [OII]$\lambda
3727$ not be detected. [OII]$\lambda 3727$ is used because it is
always apparent even in very weak emission line galaxies. Due to the
presence of complex absorption near [OII]$\lambda 3727$, we measure
this line after subtracting a preliminary PCA sample solely with the
requirement that EW($\Ha$)$<$0. A Gaussian function is fit to the
residuals between $\lambda=3700$ and $\lambda=3754$. If the $\chi^2$
of the fit is less than the $\chi^2$ of a linear fit minus 3, the line
is considered significant, and the object is rejected as an
absorption-line galaxy.

\begin{figure}[t]
\centerline{\includegraphics[angle=-90, width=\hsize]{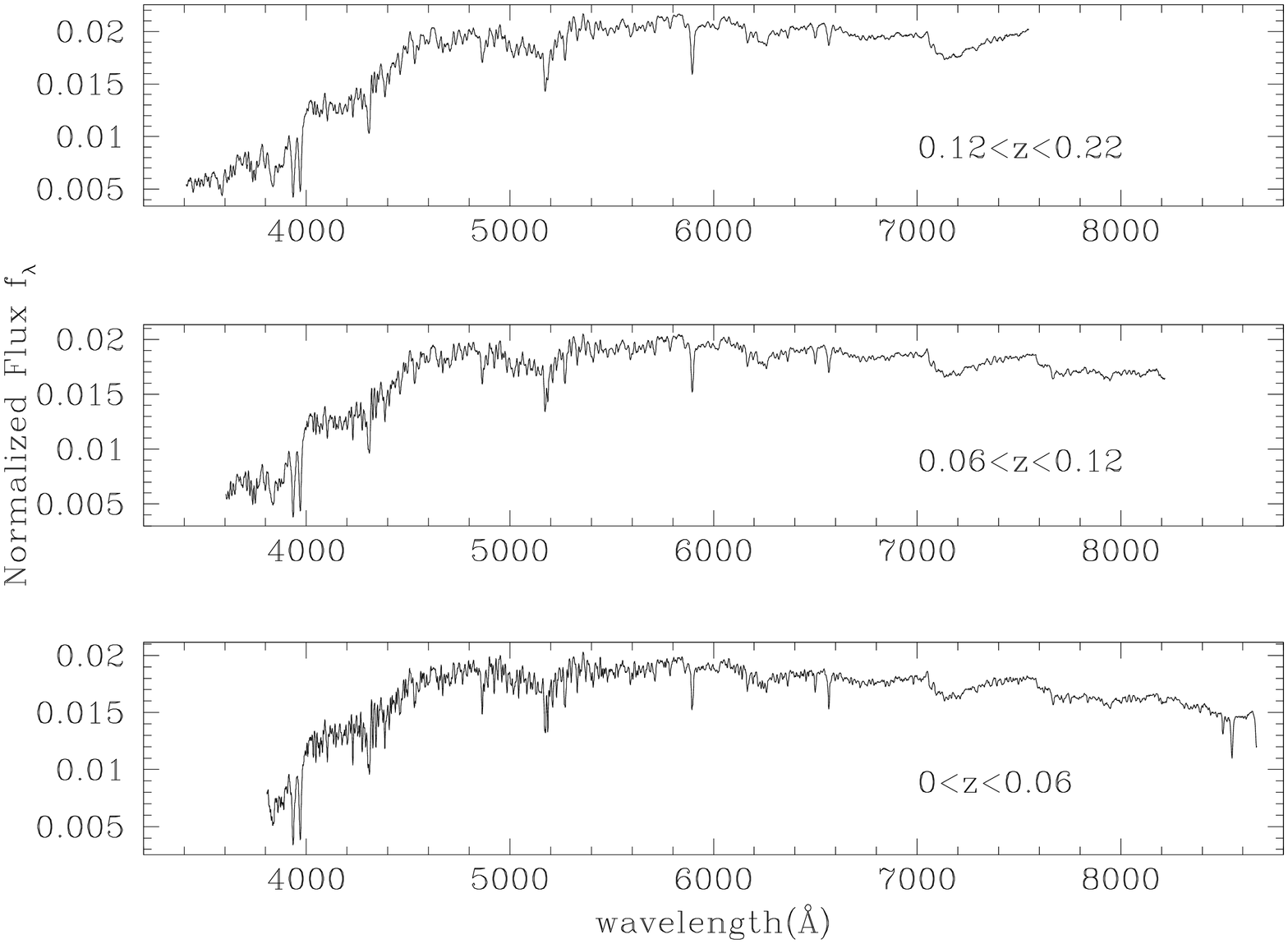}}
\caption{The mean spectra of pure absorption-line galaxies in three groups of redshift. Within the common wavelength range, the mean spectra are almost identical, although the S/N is highest at lowest redshift. PCA is applied to each group and three sets of eigenspectra are obtained.}
\label{fig:3mean}
\end{figure}

By limiting ourselves further to high S/N spectra, we defined three
samples of several hundred pure absorption line galaxies grouped by
their redshifts: 325 galaxies with $0.02<z<0.06$; 338 with
$0.06<z<0.12$ and 372 with $0.12<z<0.22$. PCA is done on each group
separately, giving three sets of eigenspectra that each can be used to
subtract stellar components for galaxies of similar redshifts. We
divide the sample in three groups in order to obtain a larger
wavelength coverage for each group as well as the resultant
eigenspectra. Figure~\ref{fig:3mean} shows the mean of the spectra
(i.e., the first eigenspectrum) in each galaxy group. They are
essentially identical, although the lowest-redshift sample has
slightly higher S/N.

The galaxies within the three groups are shifted to fixed rest-frame
wavelength bins using sinc interpolation. Afterwards, each spectrum is
normalized to a constant flux value. Unlike some PCA analyses (e.g.,
Eisenstein \etal 2003), the continua are not subtracted from the
spectra before PCA. Each sample includes several hundred normalized
galaxy spectra, which we express as a matrix $S$ of dimensions
$N\times M$, where $N$ is the total number of galaxies in each group
and $M$ is the total number of common wavelength bins. Singular Value
Decomposition (SVD) is used to build the eigenspectra (Connolly \etal
1995; Bromley \etal 1998).

\subsection{PCA Result and Stellar Subtraction}
The PCA analysis generates a set of eigenspectra, with main features
of the absorption-line galaxies concentrated in the first few. In this
study we will use the first eight eigenspectra as the stellar
absorption-line templates. Empirical tests show that including more eigenspectra, which are basically just noise, does not improve the subtraction further. A $\chi^2$ minimizing algorithm was
adopted to determine the synthetic stellar absorption spectrum for
each galaxy. The minimizing is done over the entire observed
wavelength range except regions around the strongest emission lines
found in AGN and starforming galaxies: $\Ha$, H$\beta$,
[NII]$\lambda\lambda6548,84$, [OIII]$\lambda5007$,$\lambda4959$ and
[OII]$\lambda3727$.

\begin{figure}[t]
\centerline{\includegraphics[angle=-90, width=\hsize]{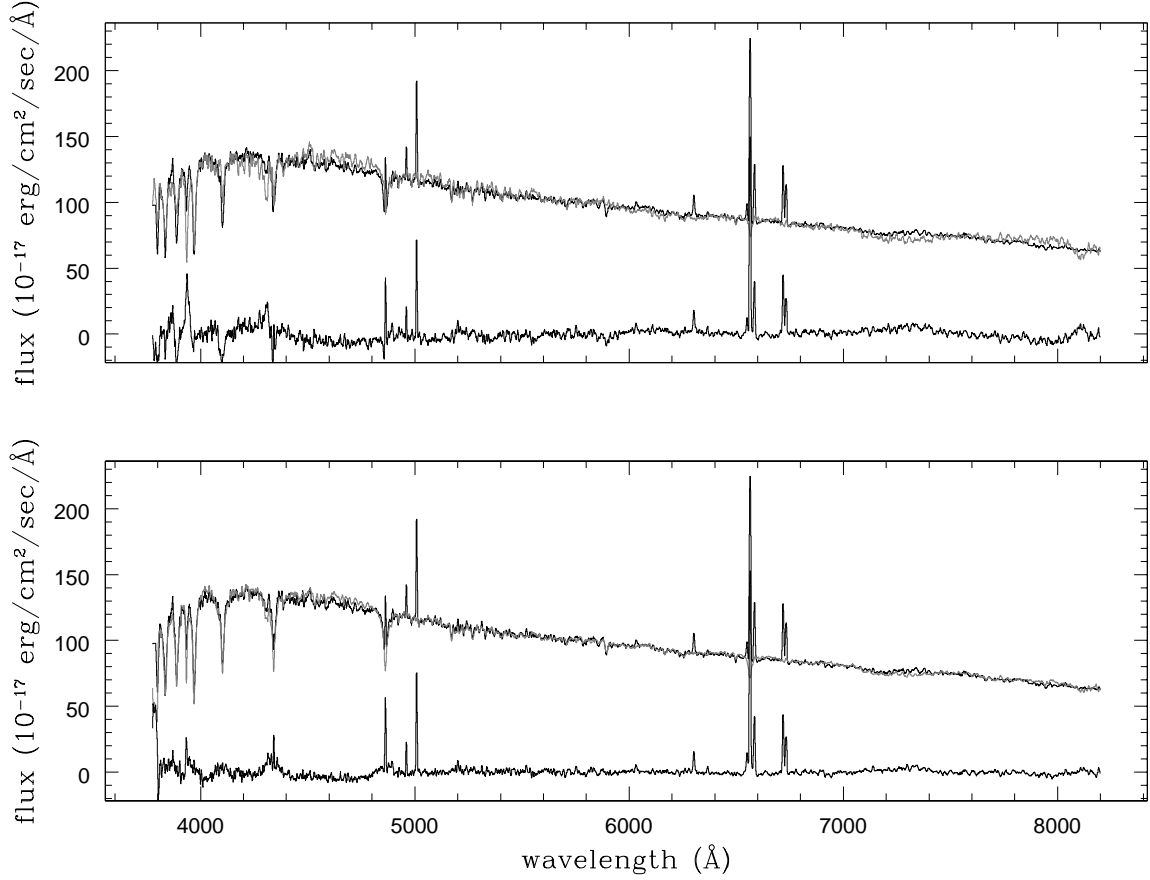}}
\caption{A subtraction example (a normal starforming galaxy) comparing
stellar-subtraction without (top) and with (bottom) an A star spectrum
included among the absorption-line templates. The grey spectra are the stellar spectra constructed using the absorption-line templates. The A star spectrum significantly improves the performance of the stellar subtraction in this case. Note that the subtraction makes weak emission lines such as [OI]$\lambda6363$ more prominent. }
\label{fig:astar}
\end{figure}

Since the stellar templates are eigenspectra of a sample of pure
absorption line galaxies and galaxies having young stellar populations
tend to have emission lines, the resultant eigenspectra mainly
represent old stellar spectral features. Thus these eigenspectra are
not representative of galaxies containing a young stellar
component. One possible resolution would be to make sure that the PCA
sample included enough E+A galaxies, which contain young stellar
populations and do not have emission lines (Goto \etal 2003; Quintero
\etal 2004). In this work, however, we simply add an A star spectrum
selected from the SDSS spectroscopic data to the absorption-line
template library to represent the young stellar
population. Figure~\ref{fig:astar} shows an example of the stellar
subtraction with and without this template, for a galaxy with a young
star population. The improvement using the A star template is
dramatic.

\begin{figure}[t]
\centerline{\includegraphics[angle=-90, width=\hsize]{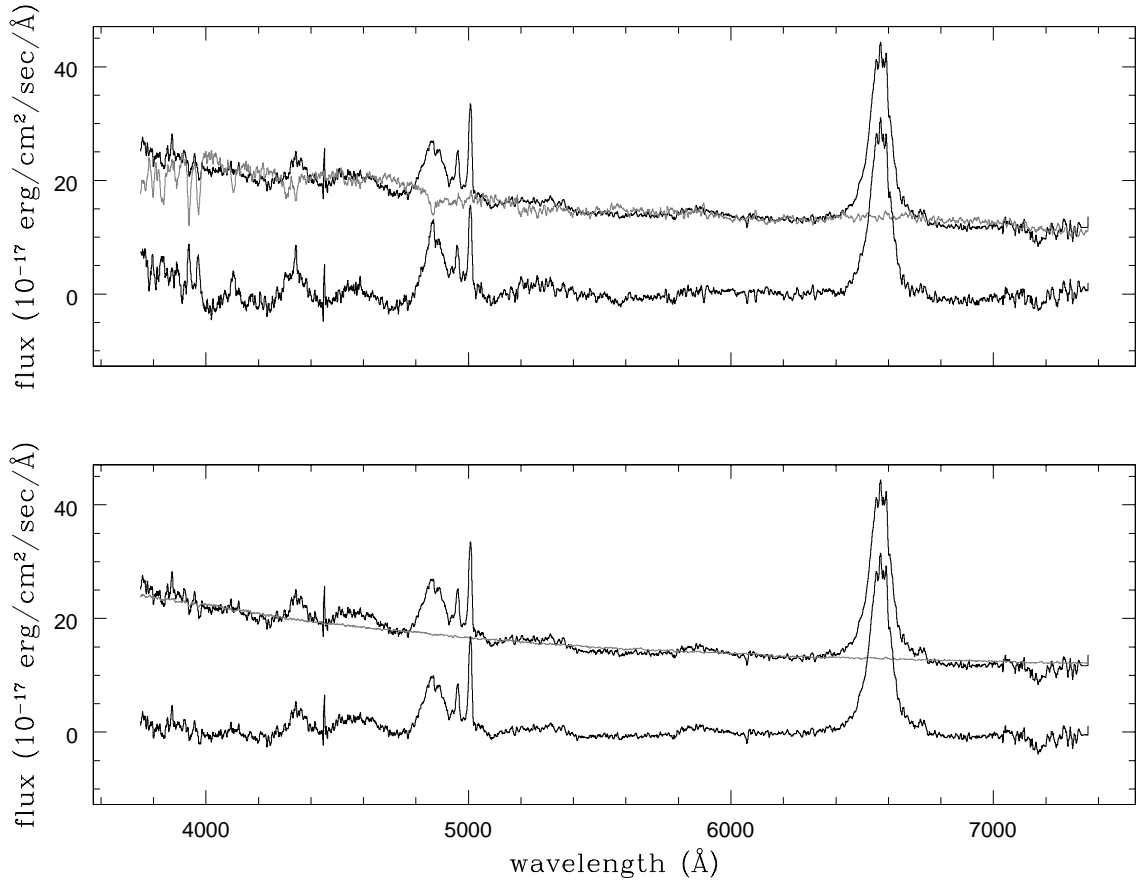}}
\caption{A subtraction example comparing stellar-subtraction without (top) and with (bottom) a power-law spectrum included in the absorption-line templates. The power-law spectrum helps to properly subtract the quasar continuum without generating fake emission lines. Note the presence of FeII $\lambda 4570$ here, which does affect the continuum fit.}
\label{fig:powerlawsub}
\end{figure}

The stellar subtraction is automatically done to all galaxies in our
parent galaxy sample, including those quasars that satisfy the galaxy
target selection magnitude cut ($\S$ 2 and 3). However, doing stellar
subtraction to a bright quasar dominated by a non-thermal continuum in
the spectrum will certainly be a disaster. We thus add a pure
power-law spectrum, written as $e_{\mathrm{powerlaw}}\propto
\lambda^{-1.5}$, to our template library. Figure~\ref{fig:powerlawsub}
shows an example of stellar subtraction for a quasar with and without
the power-law template. The power-law template significantly improves
the subtraction. Since the spectrophotometry is not perfect, quasars
have a range of intrinsic power-law slopes (Richards \etal 2002), and
the power law fit can be systematically in error due to FeII emission
(Figure~\ref{fig:powerlawsub} and Vanden Berk \etal 2001), the
power-law template will certainly not be sufficient for clean
continuum subtraction for all quasars. However we have found it to be
adequate for our purposes. For galaxies that do not have a nonthermal
power-law component, this template can help compensate for continuum
shape errors due to errors in spectrophotometry and internal
reddening.

\begin{figure}[t]
\centerline{\includegraphics[angle=-90, width=\hsize]{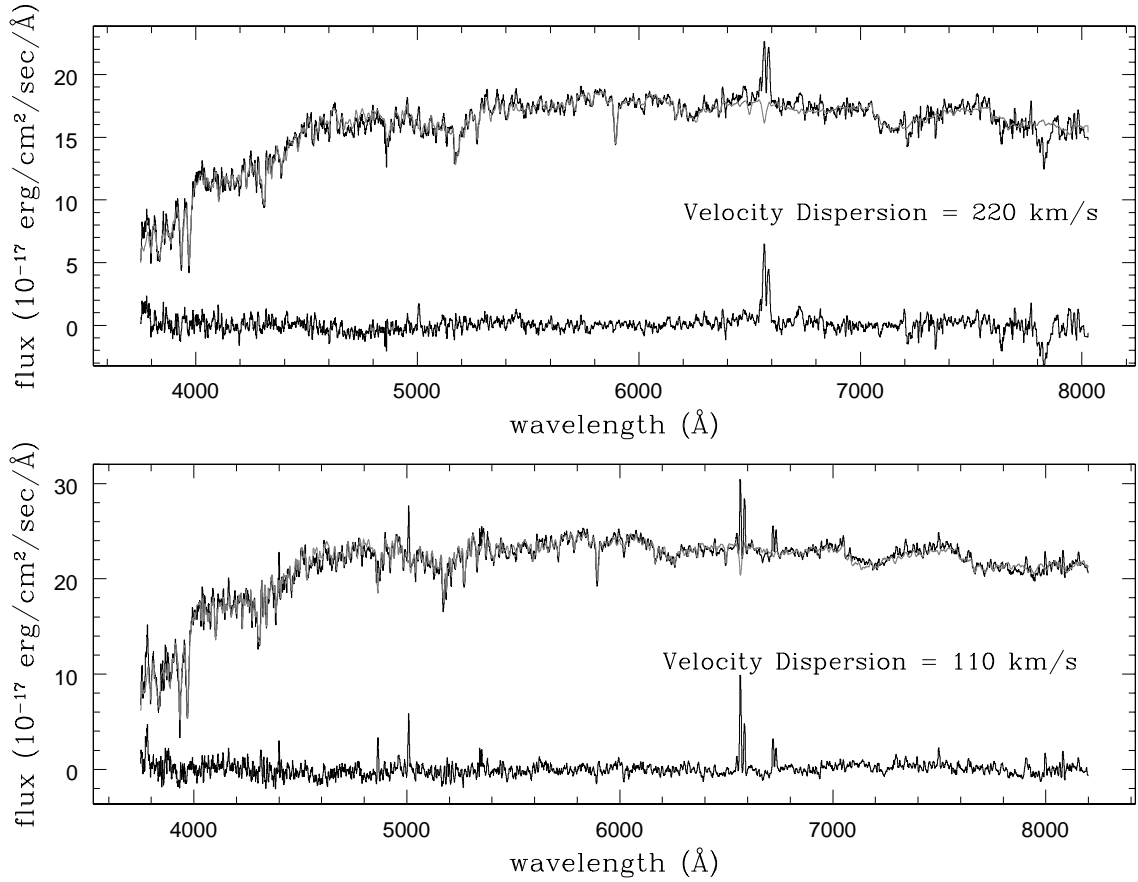}}
\caption{The subtraction using the same set of eigenspectra templates works well
for galaxies with very different host stellar velocity dispersions. The color notation is the same as in Figure~\ref{fig:astar}. Note in particular that there are no substantial residuals around the strong absorption lines in the two cases.}
\label{fig:subVD}
\end{figure}

In summary, our template library includes 8 PCA eigenspectra of pure
absorption line galaxies, a power-law continuum and an A star
spectrum. As demonstrated in Figures~\ref{fig:astar},
~\ref{fig:powerlawsub} and further in Figure~\ref{fig:subVD}, weak
emission lines such as H$\beta$ that were originally submerged in the
stellar absorption line are successfully recovered. Moreover, emission
lines such as [OI]$\lambda6363$ and [NI]$\lambda5200$ which are not
apparent at all in the original spectrum clearly stand out after the
subtraction. The subtraction also helps correct the strength of $\Ha$,
which is very important in subsequent AGN identification using
emission line ratios.
 
The stellar subtraction is robust for galaxies of a range of velocity
dispersions.  Figure~\ref{fig:subVD} shows two galaxies with very
different velocity dispersions (Bernardi \etal 2003); note that in
both cases, the subtracted spectrum shows no appreciable residuals of
the strong absorption lines. The eigenspectra include terms that can
give absorption lines of different width. This is another advantage of
the PCA technique over the use of stellar libraries: we do not need to
convolve each template with a Gaussian broadening function for each
galaxy.

The whole procedure of doing PCA and stellar subtraction using the
resultant eigenspectra is quite straightforward. Once the stellar
component is subtracted, the emission lines can be measured, as we now
describe.

\section{Emission Line Measurements}
Since we are mainly interested in emission-line galaxies, we will
first set up a criterion to remove the pure absorption-line galaxies
from the parent sample. We require that the EW of the $\Ha$ line (in
the rest frame) be greater than $3 \Angstrom$. In the rare cases in
which the $\Ha$ line is saturated or affected by bad pixels, we
examine the equivalent width of the [OIII]$\lambda 5007$ and H$\beta$
lines, requiring that one of them be greater than $3\Angstrom$. These
equivalent widths are based on the line strength after stellar
continuum subtraction, divided by the stellar continuum itself.

This criterion also rejects weak emission line galaxies. This is
acceptable since the S/N ratio for these lines will be low and thus
our ability to distinguish AGN from starbursts will be poor. A total
of 42,435 galaxies, about half of the parent galaxy sample, pass the
equivalent width cut. We refer to this sample as the emission line
galaxy sample. All AGN are selected from this sample.

The intensity, full width at half maximum (FWHM), central wavelength
and nearby continuum value of the main emission lines in these
galaxies are measured via Gaussian fits weighted by the estimated
errors per pixel. The following emission lines are needed for AGN
selection: H$\alpha$, [NII]$\lambda 6584,48$, H$\beta$, [OIII]$\lambda
5007$, [SII]$\lambda 6716,31$ and [OI]$\lambda 6300$. The following
lines are fit with a single Gaussian: H$\beta$, [OIII]$\lambda 5007$
and [OI]$\lambda 6300$. The [SII] doublet is fit with a two-Gaussian
function model. The FWHMs and intensities of the two Gaussians are
independent of each other, while the central wavelengths are
correlated with a single variable $z$, and the continuum is shared by
the two Gaussian functions. Since our fitting is done to a stellar
continuum subtracted spectrum that has been shifted to rest
wavelength, the fitting results for $z$ and the continuum are very
close to zero. The H$\alpha$ and [NII]$\lambda 6548,84$ lines are
fitted with three Gaussians. The FWHMs of the [NII] lines are kept the
same and the intensity ratio of [NII]$\lambda 6584$ to [NII]$\lambda
6548$ is fixed to 3, as required by the energy level structure of the
[NII] ion (Osterbrock 1989). Again, the central wavelengths of $\Ha$
and the [NII] doublet are correlated with a single redshift parameter
and the three lines share the same continuum value. The pixels located
within $100\Angstrom$ of the central wavelength of the emission lines
are used for Gaussian fits (adjacent emission lines in this range are
shielded out). However, in order to be sensitive to a broad component
of $\Ha$, we fit the $\Ha$, [NII] group to the range $6565\Angstrom
\pm 300\Angstrom$ (the adjacent [OI] and [SII] lines are masked in the
fit). We will use the $\chi^2$ of the fit to test for the significance
of the broad component, restricting ourselves to the range
$6565\Angstrom \pm 80\Angstrom$ to reduce the sensitivity to
uncertainties in the continuum.

Some AGN show both broad and narrow permitted emission lines. For
these galaxies, a single Gaussian function for $\Ha$ is obviously not
appropriate. To take this into account, we also fit the $\Ha$ and
[NII] doublet with a four-Gaussian model: two for the [NII] lines and
two for $\Ha$. The two $\Ha$ Gaussian functions have the same central
wavelength, but different intensities and FWHMs. The final decision of
which model (the three-Gaussian model or the four-Gaussian model) to
use for a given galaxy is done by comparing the $\chi^2$ of the two
model fits, $\chi^2_3$ and $\chi^2_4$ respectively. In particular, we
choose the four-Gaussian model fit when:
\begin{equation} 
(\chi_3^2 - \chi_4^2 -2 )/\chi_4^2 \,>\, 0.2 
\label{eq:chi}
\end{equation}
This criterion is empirical, but is inspired by a similar statistic
for linear fitting models (Lupton 1993). If we would like to fit a
data set ($x_i, y_i$) with a linear fitting model $y_{model}=y(x)$,
and the errors $\sigma_i$ are Gaussian, then
\begin{equation}
\chi^2 = \sum_i \left( {y_i - y_{\mathrm{model}}} \over {\sigma_i} \right)^2
\end{equation} 
follows a $\chi^2_{n-k}$ distribution, where $n$ and $k$ are the
numbers of the data points and the parameters used in the model
respectively. If there are two linear models $M$ and $N$, with $k$ and
$(k-r)$ parameters respectively (i.e. model $M$ has $r$ extra
parameters), their $\chi^2$ functions $\chi_N^2$, $\chi_M^2$ will then
follow $\chi^2_{n-k}$ and $\chi^2_{n-(k-r)}$ distributions. It can be
proved that for linear models, $\chi_M^2$ and $\chi_M^2-\chi_N^2$ are
independent, thus the quantity
\begin{equation}
f\equiv {{(\chi_N^2-\chi_M^2)/r} \over {\chi_M^2/(n-k)}}
\end{equation}
follows a $F_{r,n-k}$ distribution.  If $f$ is large, then we can
accept the hypothesis that the added parameters significantly improve
the fit.

In our case, our models are non-linear in the parameters, and
$\chi_M^2$ and $\chi_M^2-\chi_N^2$ will not be independent. The
four-Gaussian model has two more degrees of freedom than does the
three-Gaussian model, so the numerator of the criterion is written as
$(\chi_3^2 - \chi_4^2 -2 )$. The limit ``0.2'' in
equation~(\ref{eq:chi}) is an empirical number and is demonstrated to
be appropriate from our manual inspection.

However it is not always unambiguous to identify the broad $\Ha$
component using the above criterion, since not all emission lines are
well-fit with Gaussians (Strateva \etal 2003). In particular, narrow
emission lines generally have extended wings at their bases (Ho \etal
1997b). If the narrow emission line is strong, this non-Gaussian
feature will become prominent, and a 4-Gaussian model will be chosen
by Equation~\ref{eq:chi}. In this case, the height of the broader
component $h_1$ will be small compared to the height of the narrow
component, $h_2$ and the Gaussian width of the broader component
$\sigma$ will be relatively small. To not count such cases as broad
line, we stick with the 3-Gaussian fit for those objects which
satisfy,
\begin{equation}
\sigma<20\Angstrom (\sim 2200 {\rm km\;s}^{-1}) \;\; \;\;{\rm and } \;\; \;\;h_1/h_2< 0.1 
\label{eq:chicorrect}
\end{equation}
no matter what the criterion of equation~(\ref{eq:chi}) might indicate. 

We compared our empirical criterion (equation~(\ref{eq:chicorrect}))
with the well defined Bayesian Information Criterion (BIC, Liddle
2004). We found that we are as efficient in choosing broad-line
component as BIC, and at the same time, our criterion 
identifies far fewer fake broad-line components than BIC does.

After fitting all relevant emission lines, the line strengths are
calculated based on the line fitting parameters. The observed SDSS
spectra are accompanied by an estimated error per pixel, based on
photon and read noise statistics and variation among sky
spectra. These errors are approximately independent between pixels
and are good to 8\% (McDonald \etal 2004). In stellar continuum
subtraction we assumed the stellar templates perfect since they
are very high S/N. The errors in the line strengths are calculated
using standard propagation of errors. Those emission lines that are
less than $3\sigma$ detections are marked as weak. If they are needed
in identifying an AGN, special care should be applied ($\S 6.2$).

\section{AGN Statistics}
\subsection{Broad-Line AGN}
\begin{figure}[t]
\centerline{\includegraphics[angle=0, width=\hsize]{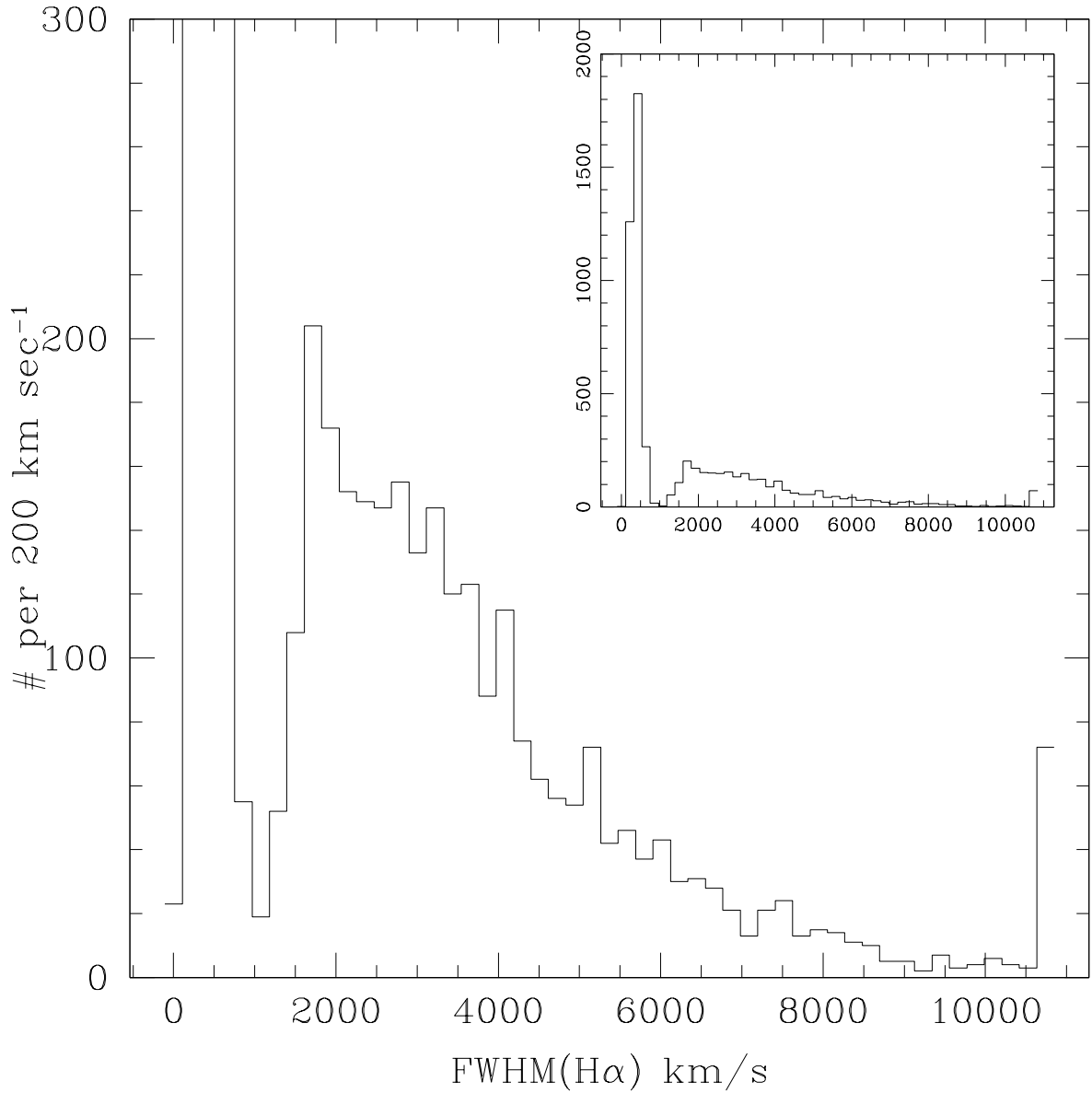}}
\caption{The $\Ha$ FWHM distribution for emission line galaxies. The inserted plot is the $\Ha$ FWHM distribution for narrow-line AGN and broad-line AGN after removing star-forming galaxies. }
\label{fig:broad}
\end{figure}

The measured emission line parameters can be used to identify AGN from
the emission line galaxy sample. We select broad-line AGN by checking
the $\Ha$ line width. Figure~\ref{fig:broad} gives the distribution of
the FWHM values of the $\Ha$ emission line for over 40,000 emission
line galaxies. For galaxies that prefer two Gaussian components for
$\Ha$ ($\S 5$), the width of the broader component is plotted. The
distribution is clearly bimodal, with a minimum at 1,200 km
s$^{-1}$. The first peak is at about 200 km s$^{-1}$, which reflects
the resolution limit of the SDSS spectrographs. It extends far above
the limit of the plot, demonstrating that most of the galaxies in the
sample are starforming galaxies with narrow $\Ha$ lines. Clearly
separated from these normal galaxies, the second group of galaxies has
substantially larger $\Ha$ FWHM value. Naturally, we use
FWHM($\Ha$)$>1,200$km s$^{-1}$ as the selection criterion for defining
broad-line AGN.
 
The inserted plot in Figure~\ref{fig:broad} shows the $\Ha$ FWHM
distribution for AGN only (the detailed selection of narrow-line AGN
is described in $\S 6.2$). Narrow-line AGN have typical $\Ha$ FWHM
values similar to those of usual emission-line galaxies. The paucity
of objects with FWHM $\sim 1,200$ km s$^{-1}$ is not well understood
and will need to be explained in AGN unification models.

\begin{figure}[t]
\centerline{\includegraphics[angle=0, width=\hsize]{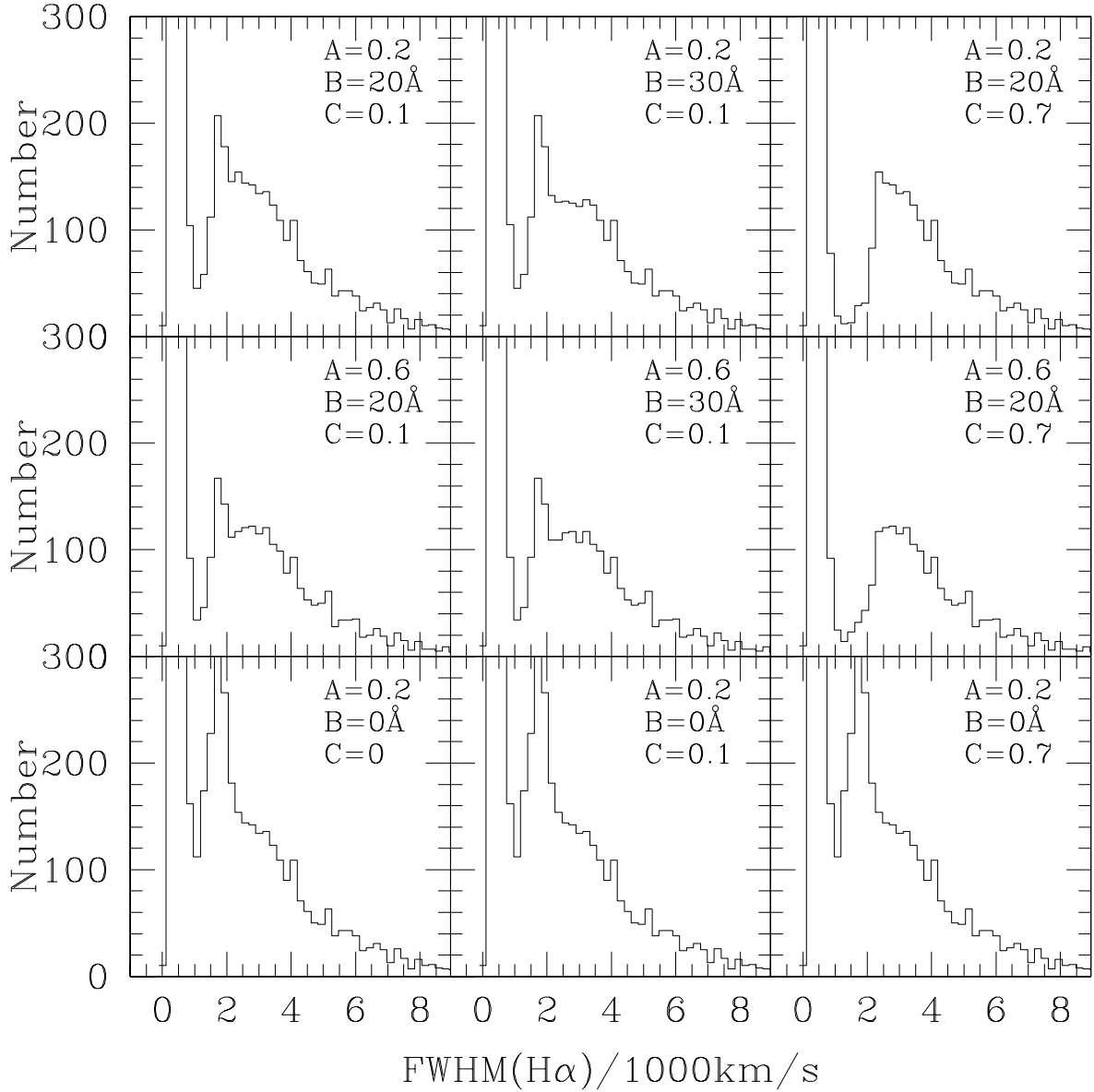}}
\caption{The distribution of $\Ha$ width for emission line galaxies with various criteria (cf, equation~(\ref{eq:4fitall_abc})) for choosing the four-Gaussian model for the $\Ha$ and [NII] lines over the three-Gaussian model. The existence of the bimodal feature is insensitive to the exact criteria used.}
\label{fig:bimocheck}
\end{figure}

However, we should make sure that the bimodality is not a consequence
of our fitting procedure in choosing the 4-Gaussian fitting model over
the 3-Gaussian model. To do so, we intentionally change the limiting
numbers of the criteria described in $\S 5$. We write the criteria
(equations~(\ref{eq:chi}) and (\ref{eq:chicorrect})) as:
\begin{equation} 
(\chi_3^2 - \chi_4^2 -2 )/\chi_4^2 \,>\, A \;\; \;\;{\rm and } \;\; \;\; (\sigma\ge B \;\; \;\;{\rm or } \;\; \;\;h_1/h_2\ge C)
\label{eq:4fitall_abc}
\end{equation}
Our default values are $A=0.2$, $B=20 \Angstrom$ (2182 km s$^{-1}$),
$C=0.1$. Now we change this set of numbers to $A=0.6$, $B=30\Angstrom$
(3249 km s$^{-1}$), $C=0.7$ separately. The changes will all reduce
the number of objects with broad $\Ha$ components. However, those
galaxies which are distinctively broad will still be selected by the
new criteria. The top two panels of Figure~\ref{fig:bimocheck} shows
the $\Ha$ FWHM distribution with the changed criteria. As can be seen,
no matter how the criteria are changed, the bimodality feature and the
minimum at about 1,200 km s$^{-1}$ remain the same even though the
number of selected broad-line AGN changes substantially. A
complementary test is to loosen the criteria a little. Out of the
concern that the $B=20 \Angstrom$ cut might artificially reject
objects with FWHM around 1,200 km s$^{-1}$, we change this criterion
to $B=0\Angstrom$. The bottom panel of Figure~\ref{fig:bimocheck}
shows the $\Ha$ FWHM distribution with $A=0.2$, $B=0\Angstrom$ and
$C=0, 0.1, 0.7$ respectively. This yields many false broad-line
AGN. However, the bimodality and the location of the minimum are still
the same. Therefore, the bimodal feature is insensitive to the
specific details of the criteria in equation~(\ref{eq:4fitall_abc}).

\begin{figure}[t]
\centerline{\includegraphics[angle=0, width=\hsize]{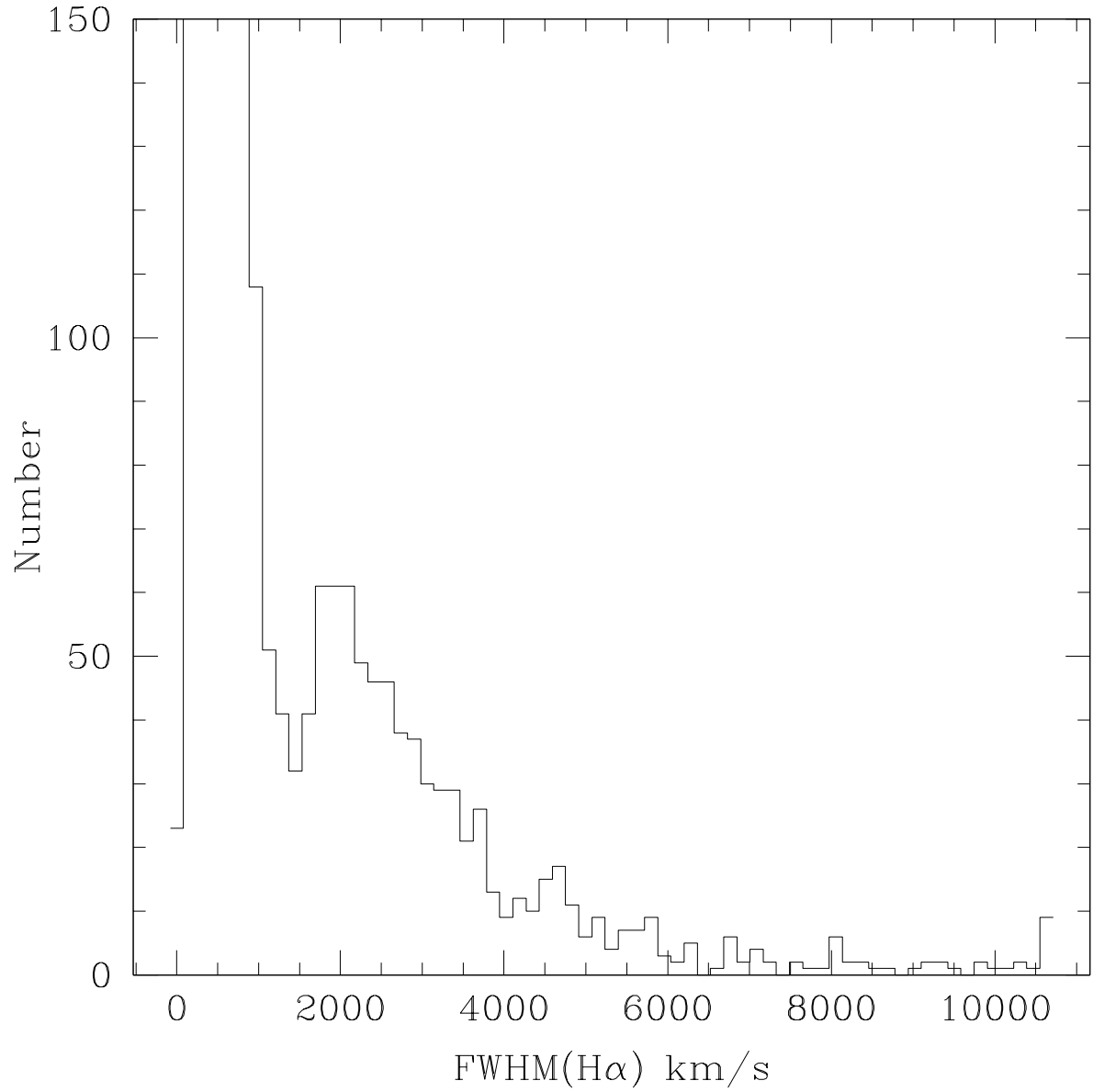}}
\caption{The $\Ha$ FWHM distribution of emission line galaxies, when the $\Ha$ and [NII] lines are fitted with a three-Gaussian model in all galaxies. }
\label{fig:3fitbimodal}
\end{figure}

One might also suspect that the bimodality is caused by the fact that
the values we have used in the plot are coming from two different
fitting models. In the distribution plot, the values of the narrower
group usually come from the fitting parameters of the 3-Gaussian model
while those of the broader group are mostly from the 4-Gaussian
model. The change from one model to another might cause some
discontinuity in an otherwise continuous distribution. To test this,
Figure~\ref{fig:3fitbimodal} shows the distribution of width using the
3-Gaussian model fits alone; the bimodal feature is still apparent. 

As a final check of the reality of the deficit of galaxies with
FWHM$\sim 1,200$ km s$^{-1}$, we carried out simulations of galaxies
with composite $\Ha$ profiles, in which the broad component had 1200
km s$^{-1}$ $\le$ FWHM $\le$ 2200 km s$^{-1}$ and $0.1 \le h_1/h_2 \le
0.3 $, and to which we add a stellar continuum and characteristic
noise. After putting these objects through our code, we found that we
would miss only a small fraction of broad-line AGN with $\Ha$ FWHM
close to 1,200 km s$^{-1}$, not nearly enough to explain the
bimodality. Therefore, the bimodal feature is clearly real.

Defining broad-line AGN as objects with $\Ha$ FWHM $>1,200$km
s$^{-1}$, 1317 broad-line AGN are identified from 42,435 emission line
galaxies. For those galaxies that need two $\Ha$ Gaussian functions
but are not classified as broad (FWHM($\Ha$)$<1,200$km s$^{-1}$), we
use a 3-Gaussian fit for $\Ha$ and [NII] so that we can have
consistent $\Ha$ measurements when they are used in emission line
ratio diagnostic diagrams, as we now describe.

\subsection{Narrow-Line AGN}

\begin{figure}[t]
\centerline{\includegraphics[angle=0, width=\hsize]{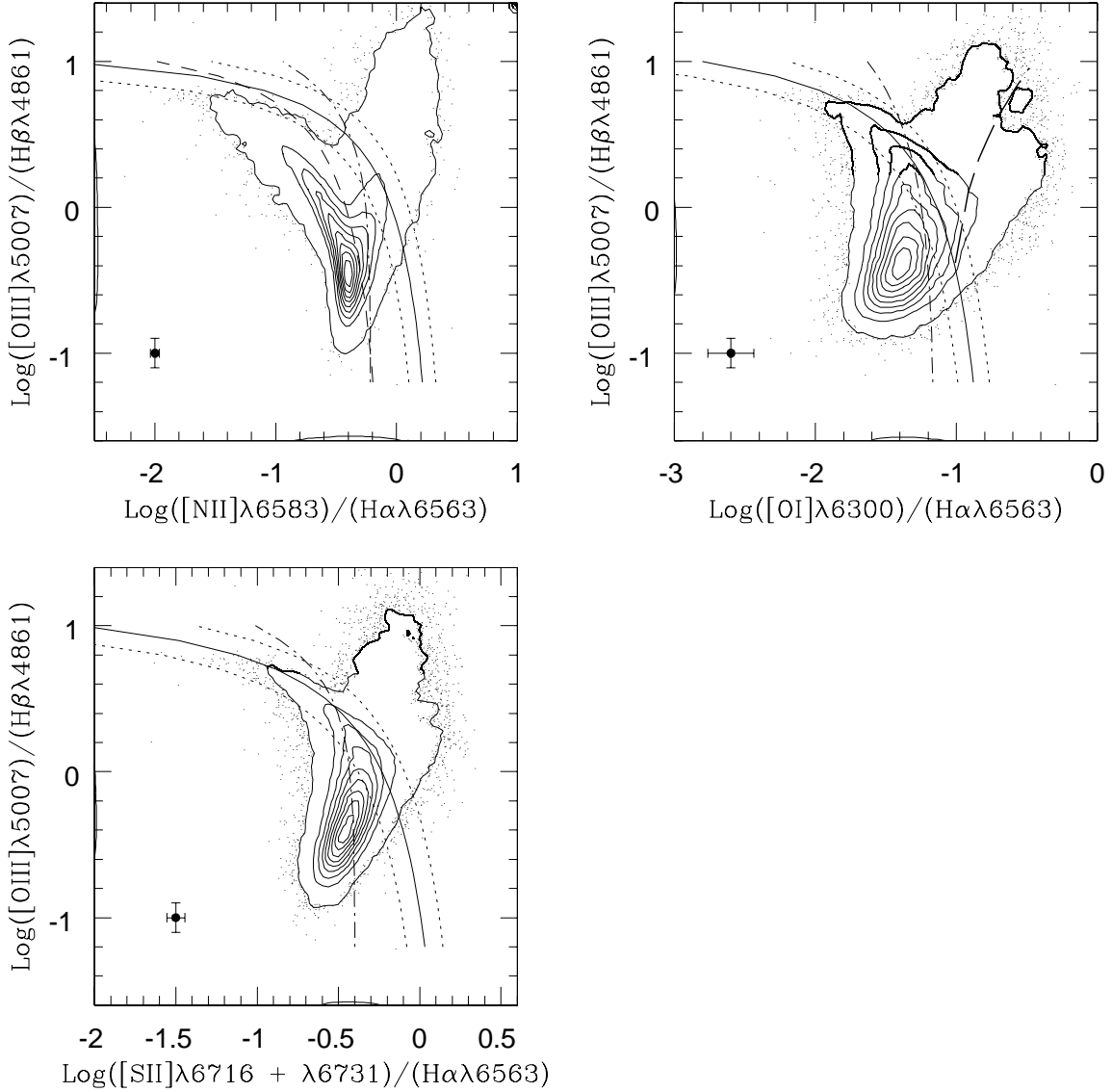}}
\caption{Emission-line diagnostic diagrams with the separation lines
taken from Kewley \etal (2001) (solid), Kauffmann \etal (2003a)
(short-dashed) and Veilleux \& Osterbrock (1987) (dot-dashed). The
dotted lines are the $\pm 0.1$ dex of Kewley's separation. The contour
plots are density contours of narrow emission line galaxies with relevant
emission lines detected at least $3\sigma$ significance (the typical
error is shown in the lower left corner). The long-dashed line in the [OI]/$\Ha$ vs. [OIII]/H$\beta$ diagram is the extreme mixing line (Kewley \etal 2001), to the right of which are possible AGN with very low ionizations, i.e. LINERs. }
\label{fig:diagram}
\end{figure}

Narrow-line AGN cannot be separated cleanly from star-forming galaxies
by their emission line widths. Therefore, we will use the traditional
emission line ratio diagnostic diagrams (Veilleux \& Osterbrock 1987;
Section 1) to select them. In Figure~\ref{fig:diagram} we place the
galaxies in the emission line galaxy sample in these
diagrams(broad-line AGN selected in Section 6.1 are excluded). Galaxies with relevant emission lines detected with less
than $3\sigma$ significance are omitted in the diagrams. There are
roughly 40,000 galaxies contained in the [NII]/$\Ha$ and [SII]/$\Ha$
diagrams, but only about 23,000 galaxies in the [OI]/$\Ha$
diagram. Due to the large number of galaxies available, we can see a
continuous distribution in the diagrams. AGN have larger
[OIII]/H$\beta$, [NII]/$\Ha$, [OI]/$\Ha$ and [SII]/$\Ha$ values than
do starbursts, and are clearly separated from the loci formed by
starforming galaxies. The AGN and starforming galaxy separation lines
developed by Kauffmann \etal (2003a, short-dashed), Kewley \etal
(2001, solid) and Veilleux \& Osterbrock (1987, dash-dot) are plotted
in the diagram. Kauffmann's and Kewley's criteria agree much better
with the shape of the density profile; the disagreement between
Kewley's lines and the data are almost all in the $\pm 0.1$dex error
range (dashed lines).

Within Kewley's separation scheme, we can identify AGN as those
located to the upper right of the lines. About 10\% of all emission
line galaxies are classified differently in the three diagrams. We
will refer to these as mismatched galaxies. One half of these are
classified as AGN by the [OI]/$\Ha$ diagram, but not by the other two.
The same discrepancy was also noticed by Stasi\'{n}ska \& Leitherer
(1996) and Dopita \etal (2000), who ascribed the enhancement of [OI]
lines in starburst galaxies to the mechanical energy released into the
gas by supernovae and stellar winds. Dopita (1997) noted that when
even a weak shock compresses the gas, the postshock local ionization
is sharply lowered, leading to a strong enhancement of the [OI]
lines. Another quarter of the mismatched galaxies are those classified
as AGN in the [SII]/$\Ha$ diagram but not in the [NII]/$\Ha$ diagram,
and the [OI] line is of too low significance to use the [OI]/$\Ha$
diagram. The cause could again be shock excitation: the relatively
cool high-density regions formed behind the shock front emit strong
[SII] emission lines (Dopita \etal 2000). The remaining quarter of the
mismatched galaxies are randomly distributed, possibly due to
noise. Overall, about 95\% of the mismatched objects have enhanced
[OI] or [SII] emission lines compared to the [NII] lines, perhaps due
to shock excitation from supernovae and stellar winds in a starburst
galaxy. The small overall number of mismatches demonstrates that
Kewley's theoretical separation lines give very consistent excitation
mechanism classifications among the three diagrams.

Using Kewley's scheme, we define the galaxies located to the upper
right of the lines in all three diagrams as narrow-line AGN. If the
emission lines used in the diagrams are weak ($\S 5$), we take special
care: if the [NII] or [OIII] line is found to be weak, the galaxy will
not be identified as an AGN unless the $\Ha$ line is also
broad. However, if the [OI] line or [SII] line is found to be weak,
then the classification based on the [OI]/$\Ha$ or [SII]/$\Ha$ diagram
is defined to be inconclusive and the identification of the galaxy
depends on the remaining diagrams. Galaxies with one or two
inconclusive classifications and AGN classifications in the remaining
diagram(s) are also included in our sample. We do not include any
mismatched objects in the final narrow-line AGN sample. Based on these
criteria, 3074 narrow-line AGN are selected.

The selection criteria are different from those used by Kauffmann
\etal (2003a) (Equation~\ref{eq:kauff}, Figure~\ref{fig:diagram}
short-dashed line). In the [NII]/$\Ha$ vs. [OIII]/$\Hb$ diagram,
galaxies are distributed on two arms. Normal galaxies have weaker
[NII]/$\Ha$ ratio than do AGN. So it is expected that they form the
locus on the left arm (with smaller [NII]/$\Ha$ values) and galaxies
on the right arm are AGN. But Kewley's line cuts through the right
arm. Kauffmann \etal propose to classify all galaxies located on the
right arm as AGN, generating an empirical line separating the two
branches. This line lies well below Kewley's criterion in the
[NII]/$\Ha$ diagram and the number of galaxies which fall between the
two criteria is very large.

\begin{figure}[t]
\centerline{\includegraphics[angle=0, width=\hsize]{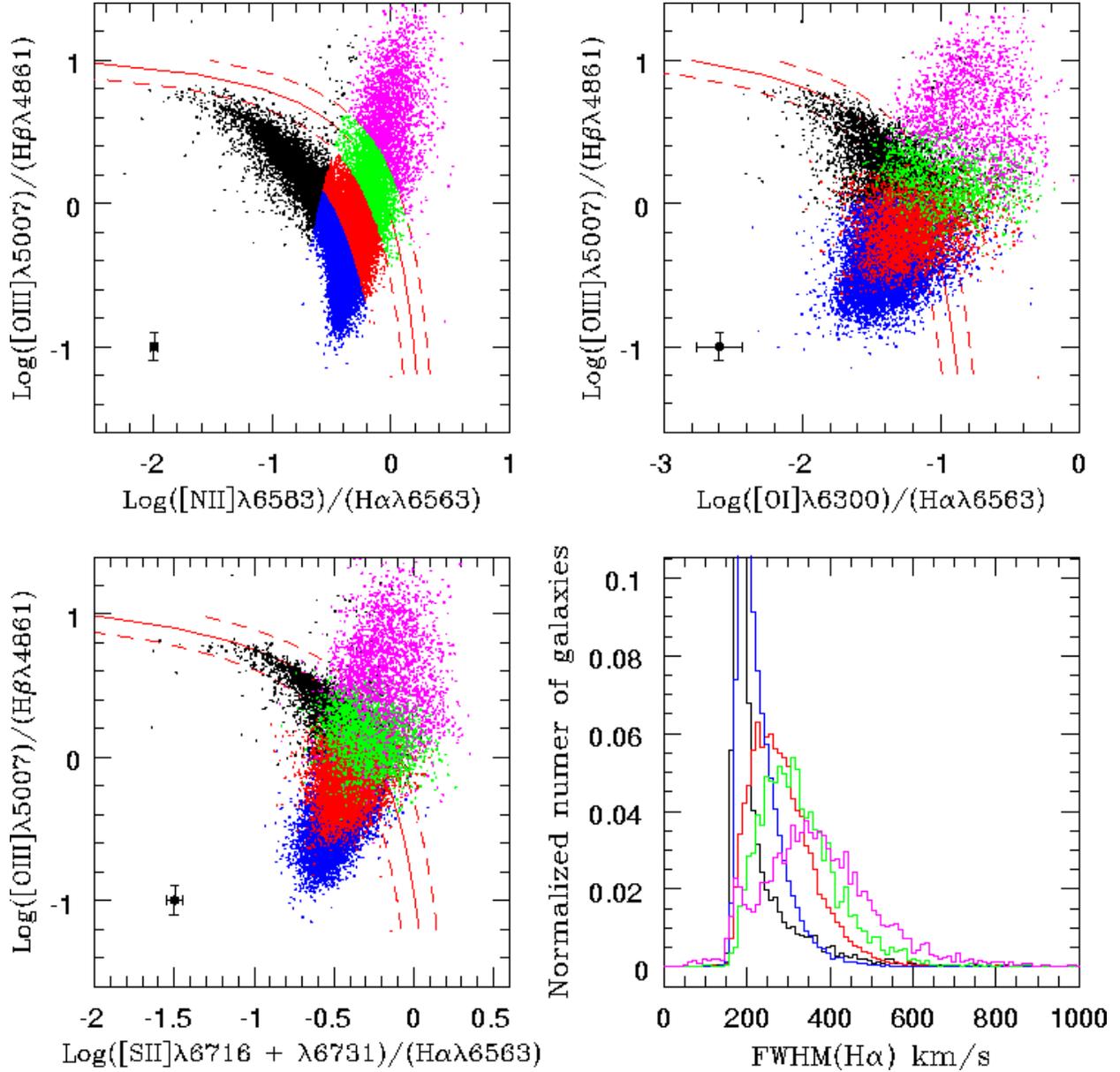}}
\caption{Diagnostic diagrams with Kewley's separation lines
(solid). The galaxies are colored by their locations on the
[NII]/$\Ha$ diagram. In the lower right figure, the distribution of
$\Ha$ FWHM of galaxies in each color group is plotted in the
corresponding color. The typical error is shown in the lower left corner.}
\label{fig:2ndarm}
\end{figure}

This difference between the two criteria will cause tremendous
disagreement on AGN statistics. If Kauffmann's criterion is adopted,
we would have selected about 10,700 narrow-line AGN, roughly three
times the AGN selected via Kewley's criterion.  To understand the
disagreement better, we divide the galaxies in the right arm into
several groups by their locations along the arm; each group is
assigned a different color in Figure~\ref{fig:2ndarm}. These galaxies
are accordingly plotted in the [OI]/$\Ha$ and [SII]/$\Ha$
vs. [OIII]/$\Hb$ diagrams. In addition, we plot the $\Ha$ FWHM
distribution for the galaxies in each group. Very interestingly,
galaxies in the right arm form a similar sequence in the three
diagrams: galaxies with larger [NII]/$\Ha$ values usually have larger
[OI]/$\Ha$ and [SII]/$\Ha$ values. Furthermore, galaxies with larger
[NII]/$\Ha$ or [OIII]/$\Hb$ values typically have broader $\Ha$
emission lines. This gives us a hint that some galaxies on the right
arm might be galaxies with mixed AGN and starburst components, with a
larger AGN component corresponding to a higher position on the
diagrams (Kauffmann \etal 2003a). However, some of the AGN and
starburst composite galaxies are degenerate with starforming galaxies
with widely spread metallicity and ionization parameters, especially
in the [OI]/$\Ha$ and [SII]/$\Ha$ vs. [OIII]/$\Hb$ diagrams. Because
of the degeneracy, Kewley's separation lines, which are the upper
limit for pure starburst galaxies, are unable to identify every galaxy
with an AGN contribution.

Therefore, the AGN sample created using Kauffmann's criterion includes
these AGN + starburst composite galaxies; they make up the majority of
the sample. While these galaxies include an appreciable contamination
from star formation in their emission line strengths, the star
formation contribution to [OIII] is relatively weak (Kauffmann \etal
2003a). The AGN sample created using Kewley's criterion includes only
those that are unambiguously AGN dominated. Both samples have their
pros and cons, so we will keep both samples. In Hao \etal (2005; paper
II), the luminosity functions are measured for both samples and the
results are compared.

Some of the narrow-line AGN have very strong low ionization lines;
these objects are ``Low Ionization Nuclear Emission Line Regions''
(``LINERs''). Ever since the first definition of the LINER class by
Heckman (1980), the excitation mechanism causing LINERs has remained
unknown. There is vigorous debate on whether or not LINERs belong to
the AGN family. Filippenko \etal (1985) showed that many LINERs have a
broad emission-line component, suggesting they are indeed AGN. Kewley
\etal (2001) established a series of starburst models and developed a
set of extreme mixing lines, below which the galaxies can only be
modeled by pure shocks without an ionizing precursor, or a power-law
ionizing radiation field with an extremely low ionization
parameter. The long-dashed lines in Figure~\ref{fig:diagram} show the
extreme mixing line in the [OI]/$\Ha$ diagram. If this line alone is
adopted here in selecting LINERs, there are 650 LINERs identified in
the narrow-line AGN sample. These objects will be studied in detail in
future work and will not be discussed further here.

\begin{figure}[t]
\centerline{\includegraphics[angle=0, width=\hsize]{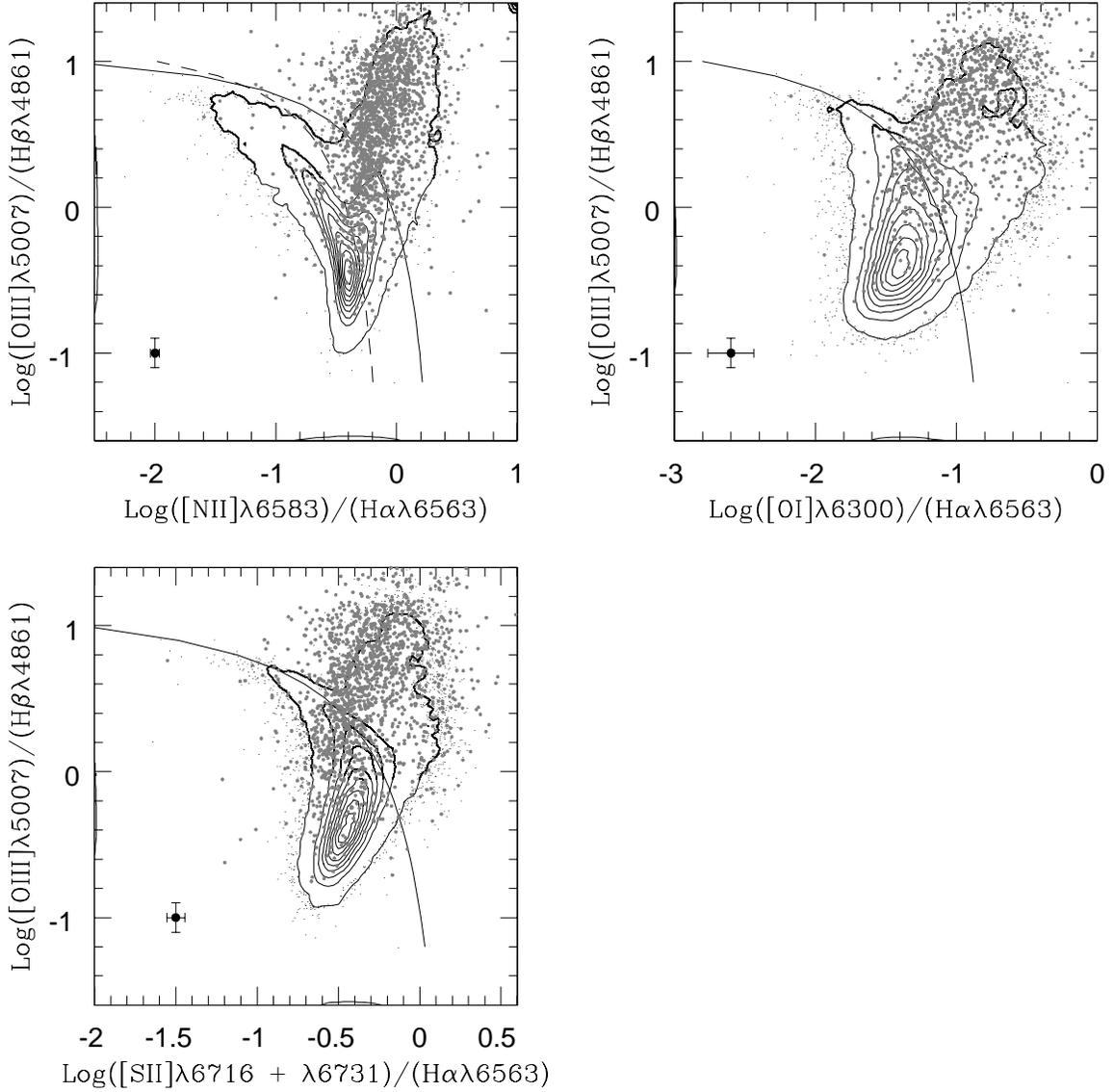}}
\caption{Diagnostic diagrams with large points representing the
locations of narrow components of broad-line AGN. The solid lines,
short dashed line and errorbars are the same as in
Figure~\ref{fig:diagram}. Note that most points are located in the AGN region of the diagram.}
\label{fig:broadindiagram}
\end{figure}

It is interesting to know how the narrow components of broad-line AGN
are located on the diagrams. In Figure~\ref{fig:broadindiagram}, we
select those broad-line AGN with real narrow $\Ha$ components (whose
line widths are less than $5\Angstrom$) and plot their emission line
ratio with the narrow components of $\Ha$ and $\Hb$ in the
diagram. Most of the broad-line AGN are concentrated on the right arm
of the [NII]/$\Ha$ vs. [OIII]/$\Hb$ diagram, indicating a strong AGN
contribution. Therefore, the classifications of these galaxies from
the narrow component and the broad component are consistent. There are
also some broad-line AGN with their narrow components corresponding to
locations to the right of the mixing line in the [OI]/$\Ha$ diagram
(Figure~\ref{fig:diagram}), these can be characterized as LINERs
(Filippenko \etal 1985).

The final AGN sample includes 1,317 broad-line AGN, 3,074 narrow-line
AGN if Kewley's criteria are adopted and 10,700 narrow-line AGN with
Kauffmann's criteria. This AGN sample is complete over 1151 square
degrees, defined by the criteria described in this paper\footnote{The
sample is available at:
\url{http://isc.astro.cornell.edu/\~{ }haol/agn/agncatalogue.txt}}.

\section{Summary}
The large number of high quality spectra available in the SDSS
provides us with great opportunities for AGN studies. In this paper,
we have systematically identified AGN from the SDSS spectroscopic data
from their emission line characteristics. In order to make unbiased
measurements of the emission lines, we have carefully subtracted
absorption stellar spectra from the observed spectra. The absorption
line templates are constructed by applying PCA to pure absorption line
galaxies. Emission lines are measured from corrected emission line
spectra by fitting Gaussian functions. In particular, to obtain proper
identification of broad emission lines, the $\Ha$ and [NII] emission
lines are fitted with both a three-Gaussian model and a four-Gaussian
model. The final decision of which model to use is based on a
comparison of $\chi ^2$ values of the two model fits.

Using measured parameters of the $\Ha$ emission lines, broad-line AGN
are identified. It is found that the distribution of $\Ha$ FWHM values
of emission-line galaxies in the parent sample is bimodal, with two
populations separated at about 1,200km s$^{-1}$. Various tests have
been applied to confirm that this feature is not caused by selection
effects. The bimodal feature also gives a natural separation between
broad-line and narrow-line AGN, and is significant in understanding
the ``unified theory'' of AGN. A naive explanation could be that the
physical parameters of the torus are correlated with the black hole
mass. For an AGN with a black hole mass less than a certain value, the
dust torus becomes so large that it almost totally blocks the
broad-line region. Therefore, it is hard to observe any broad-line
component with small $\Ha$ FWHM values.

Narrow-line AGN are identified using traditional diagnostic
diagrams. Due to the large number of galaxies available, starforming
galaxies follow well-defined loci. Theoretical separation lines
between AGN and starforming galaxies developed by Kewley \etal (2001)
agree fairly well with the shape of the galaxy density contour and
bring few mismatched classifications among the three diagrams. Only in
the [NII]/$\Ha$ diagram does Kewley's line seem to disagree with the
data, where the galaxies are distributed into two arms and Kewley's
line cuts through the right arm (larger [NII]/$\Ha$ value). Kauffmann
\etal (2003a) argued that all objects in the right arm should be
considered as AGN. Based on the SDSS data, they proposed to use the
empirical line separating the two arms to select AGN. This line is
lower than Kewley's line, and would classify many more objects than
Kewley's criterion as narrow-line AGN. We demonstrated that AGN
selected via Kewley's criterion only include those that are dominated
by active nuclear activity, while AGN selected via Kauffmann's
criterion also includes AGN + starburst composite galaxies. Both
samples have their pros and cons, and we keep both for further
studies.

The final AGN sample, complete to magnitude cut
$r(\mathrm{Petrosian})<17.77$, includes 1317 broad-line and 3074
narrow-line AGN (10,700 if Kauffmann's criterion is used) over 1151
square degrees. This is based on 1/8 the eventual SDSS survey; the
sample will grow substantially. This sample is very useful for various
studies. First, with the large number of AGN in this sample, we can
evaluate the AGN luminosity functions; we carry out this analysis in
Paper II. Second, by correlating these AGN with their host galaxy
parameters, we can study the relationship between AGN and their host
galaxies for both narrow-line (Kauffmann \etal 2003a) and broad-line
AGN. Third, from the velocity dispersion of the AGN host galaxies, we
can directly measure the approximate black hole mass using the $M_{\rm
BH}-\sigma$ relation developed first in quiescent galaxies (Gebhardt
\etal 2000; Ferrarese \& Merritt 2000; Ferrarese \etal 2001; Tremaine
\etal 2002), and further derive the accretion rate for various types
of AGN (Heckman \etal 2004). This will help us to understand their
accretion mechanisms. Fourth, we can measure the cluster properties
and local environments of different types of AGN in the sample (Miller
\etal 2003), which is important to constrain AGN formation
scenario. Furthermore, we can also correlate this AGN sample with
surveys in other wavebands, such as X-ray, radio, infrared, etc. to
better understand the physics of AGN. Studies on these topics will be
covered in future papers.

{\bf Acknowledgments.} \,\,\,We would like to thank Lisa Kewley for
useful discussions and comments.

Funding for the creation and distribution of the SDSS Archive has been
provided by the Alfred P. Sloan Foundation, the Participating
Institutions, the National Aeronautics and Space Administration, the
National Science Foundation, the U.S. Department of Energy, the
Japanese Monbukagakusho, and the Max Planck Society. The SDSS Web site
is http://www.sdss.org/.

The SDSS is managed by the Astrophysical Research Consortium (ARC) for
the Participating Institutions. The Participating Institutions are The
University of Chicago, Fermilab, the Institute for Advanced Study, the
Japan Participation Group, The Johns Hopkins University, the Korean
Scientist Group, Los Alamos National Laboratory, the
Max-Planck-Institute for Astronomy (MPIA), the Max-Planck-Institute
for Astrophysics (MPA), New Mexico State University, University of
Pittsburgh, Princeton University, the United States Naval Observatory,
and the University of Washington.

L.H. and M.A.S. are supported in part by NSF grant AST-0071091 and
AST-0307409. L.H. acknowledges the support of the Princeton University
Research Board.

\clearpage

\clearpage

\clearpage

\clearpage

\clearpage

\clearpage

\clearpage

\clearpage

\clearpage

\clearpage

\end{document}